\documentclass[aps,prl,twocolumn,superscriptaddress,showpacs]{revtex4}
\usepackage{graphicx}

\begin{document}

\title{Adiabatic Landau-Zener-St\"uckelberg transition with or without
dissipation in low spin molecular system 
V$_{15}$}

\author{I. Chiorescu}
\altaffiliation[Present address: ]{Department of Applied Physics, Delft
University of Technology, 
2628 CJ, Delft, Netherlands}
\affiliation{Laboratoire Louis N\'eel, CNRS, BP 166, 38042-Grenoble, France}
\author{W. Wernsdorfer}
\affiliation{Laboratoire Louis N\'eel, CNRS, BP 166, 38042-Grenoble, France}
\author{A. M\"uller}
\affiliation{Fak\"ultat f\"ur Chemie, Universit\"at Bielefeld, D-33501,
Bielefeld, Germany}
\author{S. Miyashita}
\affiliation{Department of Applied Physics, University of Tokyo, Bunkyo-ku,
Tokyo, 113-8656, Japan}
\author{B. Barbara}
\affiliation{Laboratoire Louis N\'eel, CNRS, BP 166, 38042-Grenoble, France}

\date{\today}

\begin{abstract}
The spin one half molecular system V$_{15}$ shows no barrier against spin
reversal. This makes 
possible direct phonon activation between the two levels. By tuning the field
sweeping rate and the 
thermal coupling between sample and thermal reservoir we have control over the
phonon-bottleneck 
phenomena previously reported in this system. We demonstrate adiabatic motion
of  molecule spins 
in time dependent magnetic fields and with different thermal coupling to the
cryostat bath. We also 
 discuss the origin of the zero-field tunneling splitting for a half-integer
spin.
\end{abstract}

\pacs{75.50.Xx, 75.45.+j, 71.70.-d}

\maketitle


Research on quantum mechanical behavior in large systems is of fundamental
interest and also of  
major  importance in quantum computation related applications (\cite{Leggett80}
and refs. therein). 
A special class are the magnetic two-level systems, realized in mesoscopic
molecular crystals at 
miliKelvin temperatures \cite{Tupitsyn01}. Here, one scales up the quantum
properties of one 
molecule in its local environment to large enough signals generated by the
whole crystal. Any coupling 
between the \emph{spin-up} and \emph{spin-down} orientation of the molecular
spin results in 
tunneling splittings whose magnitude control the quantum dynamics of the spin.
Quantum 
tunneling of the magnetization across anisotropy barrier was observed in
Mn$_{12}$ or Fe$_{8}$ high 
spin molecules \cite{Thomas96} and quantum interference of tunneling paths was
demonstrated in 
Fe$_8$ \cite{WW_Science99}. In these molecules splittings are very small
($\lesssim10^{-6}$~K). 
Large splittings ($\gtrsim10^{-3}$~K) are usually found in low spin molecules
or high and 
intermediate spin molecules in large transverse fields 
\cite{ChiorescuPRL00b,WW_Science99,Tupitsyn01}. In such systems, the archetype
of which is the low 
spin molecule V$_{15}$  \cite{ChiorescuPRL00a,ChiorescuJMMM00,Barbara02}
Landau-Zener-St\"uckelberg 
(LZS) tunneling probability \cite{Landau32} is very close to 1 and, in the
absence of 
environment, the transitions should be adiabatic. At finite temperature,
dynamics of the magnetic 
moment depend on direct spin-phonon transitions fed by the energy exchange with
the 
cryostat. The $S=1/2$, V$_{15}$ molecule shows an intrinsic phonon bottleneck
(PB) effect with 
characteristic ``butterfly'' hysteresis loops. Such systems are experimental
realizations of 
dissipative two-level systems \cite{Leggett87}. Similar results were reproduced
since then in other 
low-spin systems \cite{Schenker02}. The PB phenomena belong to a general class
of almost adiabatic 
transitions in which only a small fraction of heat inflows during the process
of approach to 
equilibrium (which remains not completed). A similar example of almost
adiabatic transition is 
found in the Foehn effect in meteorology and similar effects (``magnetic
Foehn'' effect 
\cite{Saito01})  have been found in the magnetization processes of some
Fe-rings \cite{Nakano01}.

In this letter we present new results on mechanisms involved in the quantum
dynamics of the 
V$_{15}$ molecular spin. We prove fully adiabatical  motion of this molecular
multispin system in 
time dependent magnetic fields and give more insight on the PB effect in
showing ways to control 
dissipation in regions close to tunneling gaps. This gives the possibility to
increase both relaxation 
and decoherence times. Furthermore, we discuss the origin of the zero-field
splitting $\Delta_0$ 
for half-integer spin . 

The V$_{15}$ complex (K$_6$[V$_{15}^{\text IV}$As$_6$O$_{42}$(H$_2$O)]8H$_2$O) 
forms a lattice with  trigonal symmetry ($a=14.029$~\AA,
$\alpha=79.26^{\circ}$, $V=2632$~\AA$^3$) 
with two V$_{15}$ molecules per cell \cite{Muller88}. All the fifteen V$^{\text
IV}$ ions of spin 
$S=1/2$ are placed in a quasi-spherical layered structure formed of a triangle
sandwiched between 
two non-planar hexagons. The exchange couplings between the V ions are
antiferromagnetic making of 
V$_{15}$ a typical example of frustrated molecule. After the PB study given in 
\cite{ChiorescuPRL00a}, the next step, presented here, was to control the
relaxation process by 
tuning the coupling between the spin-phonon system and the thermal reservoir.
In order to reach a 
regime where spin dynamics is independent on the thermal bath, thus entering in
a pure quantum 
regime, we minimized the coupling to the sample reservoir. This first limit of
fast field 
variations and weak coupling to the cryostat, was reached by inserting a
thermal isolator between 
the sample (a single-crystal of few tens of microns) and the sample-holder and
restricting the 
temperature to lowest values. We optimized the other limit with best thermal
coupling by 
using a mixture of silver powder with grease. 

\begin{figure}
\begin{center}\includegraphics[width=8cm]{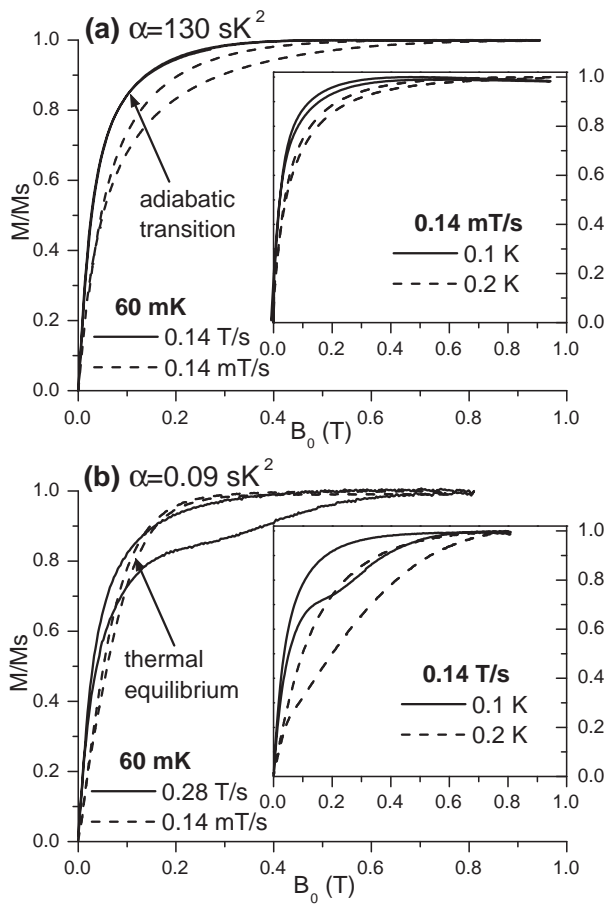} 
\end{center}
\caption{\textbf{(a)} Hysteresis loops for a sample rather well isolated from
the cryostat ($\alpha 
=130$~sK$^2$). The curve at 60~mK and 0.14~T/s shows an adiabatic LZS 
spin transition, well fitted by Eq.~\ref{eqmadiab}. \textbf{(b)} Hysteresis
loops measured in the 
case of a sample rather well coupled to the cryostat ($\alpha = 0.09$~sK$^2$).}
\label{figalpha}
\end{figure}

The magnetization curves \emph{vs.} applied field $B_0$ ($=\mu_0H$) of 
Fig.~\ref{figalpha} show either a weak hysteresis or no hysteresis at all, 
depending on the competition between two field-dependent parameters: 
relaxation rate $\tau_H$ and $\Delta_H/v_{\Delta_H}$, the time spent by system
at a given two-level 
energy separation ($\Delta_H=\sqrt{\Delta_0^2+(2\mu_BB_0)^2}$, 
$v_{\Delta_H}=\text{d}\Delta_H/\text{d}t$). In our experimental setup one may
tune the ratio 
$\tau_Hv_{\Delta_H}/\Delta_H$ roughly between $10^{-3}-10^4$ by adjusting the
field sweeping rate 
and sample thermal coupling, which makes possible to study both regimes, spin
system isolated 
(continous line, Fig.~\ref{figalpha}a) or in thermal equilibrium with the
cryostat (dashed line, 
Fig.~\ref{figalpha}b). The PB model  accounts for the interpretation
\cite{ChiorescuPRL00a} of 
such hysteresis curves (for clarity only half cycles are shown in
Figs.~\ref{figalpha}, negative 
field half cycles being symmetric to the positive field ones). The spin
temperature $T_S$ is such 
that $n_1/n_2 = \exp(\Delta_H/k_{B}T_S )$ where $n_{1,2}$ ($n_{1,2eq}$) are
levels 
out-of-equilibrium (equilibrium) occupation numbers.
Most molecules are out of equilibrium, at a temperature $T_S$ 
different from the cryostat temperature $T$. Even for a good thermal contact
between the sample and 
the cryostat, the energy flow from the latter is not sufficient to compensate
the lack of phonons 
at energies $\hbar\omega=\Delta_H\pm\Delta\omega$, if the field is swept fast
enough.  The 
phonon-bottleneck model presented in \cite{ChiorescuPRL00a,Abragam70} gives a
relaxation law: 
$-t/\tau_H=x(t)-x(0)+\ln((x(t)-1)/(x(0)-1))$ where
$x=(n_1-n_2)/(n_{1eq}-n_{2eq})$ and
\begin{equation}
\tau_H=\alpha\frac{\tanh^2(\Delta_H/2k_BT)}{\Delta_H^2},
\label{eqtauH}
\end{equation}
is the relaxation time. The $\alpha$ parameter reflects the thermal coupling
sample-cryostat (large 
value = weak coupling, small value = strong coupling); $\alpha$ is proportional
to molecule 
density, phonon velocity, phonon-bath relaxation time and inverse proportional
to levels width 
($\Delta\omega$, directly related to nuclear spin bath fluctuations
\cite{Prokofev00}, of 
the order of 50~mK \cite{Carter77}). To compare measured and calculated
hysteresis curves one has 
to find the best set of parameters $(\alpha, \Delta_0)$. The experimental
curves in 
Fig.~\ref{figalpha}a,b are best fitted by sets of parameters
($\alpha\approx130$~sK$^2$, 
$\Delta_0\approx80$~mK) and ($\alpha\approx0.09$~sK$^2$,
$\Delta_0\approx80$~mK), respectively. As 
expected, ($\alpha\approx0.15$~sK$^2$, $\Delta_0\sim50$~mK) obtained in
\cite{ChiorescuPRL00a}, 
shows a comparable $\Delta_0$ and $\alpha$ between the present ones, meaning
that the coupling with 
cryostat was intermediate compared with those of Fig.~\ref{figalpha}. It is
interesting to mention 
that recent neutron scattering experiments done on V$_{15}$ powder
\cite{Chaboussant02} obtained a 
value of $\Delta_0$ comparable to our first determinations. 

In Fig.~\ref{figalpha}b one can see that magnetization curve becomes almost
reversible if the field 
change is slow enough (dashed line). This is a natural consequence of Boltzmann
thermal equilibrium 
when spin-phonon transitions are highly probable. When sweeping rate $v_H$ or
thermal isolation 
$\alpha$ are increasing, the PB phenomena becomes more present and manifests
itself through an 
opening in the hysteresis loops. In very low fields ($\lesssim0.1$~T), the spin
temperature is lower 
than the bath temperature ($T_S < T$). Then a PB plateau of almost constant
magnetization develops 
and $T_S$ overpass $T$. At sufficiently high fields ($\sim0.5$~T) the system
reaches its 
equilibrium ($T_S = T$) by a small phonon avalanche. Of a particular interest
is the case 
$\alpha=130$~sK$^2$ when the sample is thermally isolated from the sample
holder. The reversibility 
of Fig.~\ref{figalpha}a (continous line) is very different from the equilibrium
reversibility. 
Here, the sweeping rate and isolation to the cryostat are large enough to drive
the PB plateau near the 
saturation level: the PB phenomena is so important that the sample is
completely isolated during 
our experimental time. The spin-phonon transitions have such a small
probability that the system 
keeps its initial state (here, the groundstate), allowing precise measurements
of slow (spin-phonon 
transitions free) relaxations. The multi-spin motion can therefore be
considered as adiabatic with 
an excellent approximation and magnetization curve should be given by:
\begin{equation}
M=\frac{1}{2}\frac{\text{d}\Delta_H}{\text{d}B_0}=\mu_B\left/\sqrt{1+\left(\frac{\Delta_0}{2\mu_BB_0}\right)^2}. 
\right.
\label{eqmadiab}
\end{equation}
Indeed, the adiabatic curve of Fig.~\ref{figalpha}a is nicely fitted by
Eq.~\ref{eqmadiab} with 
$\Delta_0\approx80$~mK.

\begin{figure}
\begin{center}\includegraphics[width=8cm]{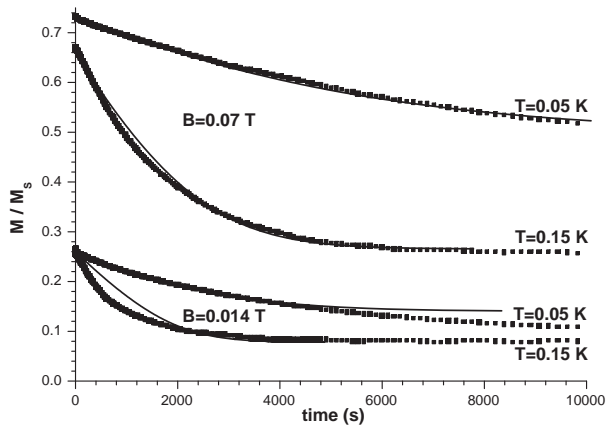} 
\end{center}
\caption{Relaxation measurements at T=0.05 and 0.15 K (dots) fitted with the PB
model (lines, see 
Eq.~\ref{eqtauH}), outside  ($B=0.07$~T) and inside the degeneracy zone
($B=0.014$~T).}
\label{figrelax}
\end{figure}

Magnetic relaxation experiments were performed in the same conditions as in
Fig.~\ref{figalpha}a, 
for different values of the applied field. As an example, we show in
Fig.~\ref{figrelax} the 
results obtained for $T=0.05$, 0.15~K and two fields especially chosen outside
and inside the 
degeneracy zone (where $\Delta_H(B_0)$ is non-linear). The corresponding
$\tau_H$ values, obtained 
by fitting the relaxation curves or simply calculated from Eq.~\ref{eqtauH}
(with 
$\alpha=130$~sK$^2$, $\Delta_0=80$~mK) are given in Tab.~\ref{tabtauH} . The
fit using the 
relaxation law mentioned above (see Eq.~\ref{eqtauH}) is in remarkably good
agreement with the data 
when the relaxation experiment is performed relatively far from the degeneracy
point. The fit is 
less performant for the faster relaxations developing inside the degeneracy
zone and the obtained 
value for $\tau_H$ is $3-5$ times shorter than the expected value (deduced from
the general shape 
of the hysteresis curve). This shows that in this case, \emph{i.e.} within the
mixing region, the 
phonons bath is no longer sufficient to explain the dissipation in this two
level system. An excess 
relaxation comes from the Dzyaloshinsky-Moriya (D-M) interactions, allowed by
molecule symmetry, 
and the spin bath, in particular from the fast fluctuations of the $^{51}$V
nuclear spins. The D-M 
interactions generates off-diagonal terms coupling the spin-up and spin-down
states, whereas the 
spin bath gives a noticeable spread of energy levels almost touching each other
near zero-field, 
both effects enhancing the relaxation in this particular region. Also, they are
at the origin of 
the large zero-field splitting $\Delta_0$ in the V$_{15}$ molecule as it will
be discussed below.
\begin{table}
\begin{center}
\begin{tabular}{|c|c|c|c|}\hline 
$T$ (K) & $B_0$ (mT) & $\tau_H$ fit (s) & $\tau_H$ th. (s) \\ \hline
0.05 & 14 & 1507 & 8716 \\
0.15 & 14 & 551 & 1323 \\
0.05 & 70 & 3883 & 3675 \\
0.15 & 70 & 970 & 997 \\
\hline
\end{tabular}
\end{center}
\caption{Relaxation times obtained by fit of curves in Fig.~\ref{figrelax}
(3$^{rd}$ column) and 
calculated (4$^{th}$ column) for $\alpha=130$~sK$^2$ and $\Delta_0=80$~mK
(obtained by fit of 
hysteresis loops Fig.~\ref{figalpha}a).}
\label{tabtauH}
\end{table}

V$_{15}$ has an effective spin $1/2$ which is a half-integer. As well known as
the Kramers theorem, 
all the energy levels have to be at least doubly degenerate in the time
reversal symmetry. 
Therefore, if we consider the tunneling of the magnetization of the
half-integer spins in a 
single mode path-integral formula, the tunneling rate must be zero. This is due
to zero tunnel 
splitting $\Delta_0$. The consequence is that magnetization reversals cannot be
adiabatic with 
half-integer spins. However, as shown above, the V$_{15}$ system shows very
clear adiabatic spin 
reversal, with zero field splitting about almost 0.1 K. The most characteristic
feature of V$_{15}$ 
is its multispin character. We shall see that this character is a necessary but
not sufficient 
condition for producing finite zero-field splittings with odd-integer spins.
Note that $\Delta_0$ 
cannot be due to internal dipolar fields which range in the mT scale. However,
these fields 
together with nuclear spins certainly provide sources of decoherence.

\begin{figure}
\begin{center}\includegraphics[width=7.97cm]{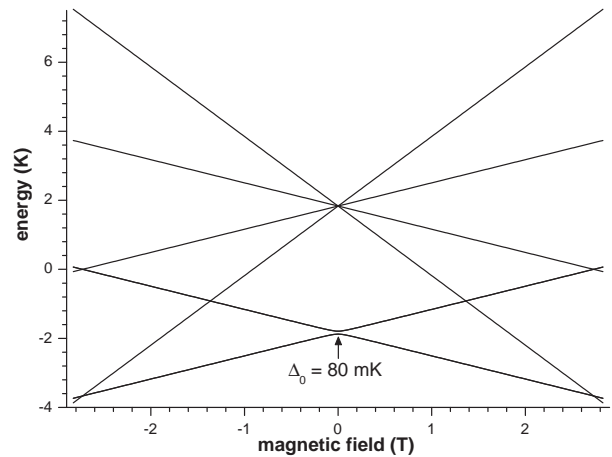} 
\end{center}
\caption{Energy levels of 3 effective spins 1/2, obtained from
$H=J\sum_{i<j}\vec S_i\cdot\vec 
S_j+D\sum_{i<j}(S_i^zS_j^x-S_i^xS_j^z)-g\mu_BB_0\sum_iS_i^z$ \cite{Miyashita02}
with $J=2.445$~K 
\cite{ChiorescuJMMM00} and $D=50$~mK (for $\Delta_0=80$~mK).}
\label{figDM}
\end{figure}

Due to frustration on antiferromagnetic triangles in V$_{15}$, the energy
structure of this system 
consists of one quartet ($S=3/2$) and two doublets ($S=1/2$) well separated
from excited levels by 
a gap of the order $10^2$ K. At zero field, the groundstate is fourfold
degenerated (although with 
a spin $S=1/2$). By making use of this degree of freedom one can have avoided
level structures 
\cite{Miyashita02}. In order to mix the states whose magnetization differs by
1, we need the
interactions to contain $S^+$ and $S^-$.  In the theoretical calculations the
transverse field is 
often used for this purpose. However, in order to find the source of the
avoided level crossing for 
the molecules in zero field, we have to find interactions with time reversal
symmetry, that is, 
products of even number of spin operators, such as $S^zS^x$, etc. Generally
symmetric combinations 
of operators, \emph{e.g.} $S^z_iS^x_j+S^x_iS^z_j$, are rearranged into
anisotropy and do not cause 
avoided level structure. This is not the case with antisymmetric combinations
of the form 
$S^z_iS^x_j - S^x_iS^z_j$ for which it has been found that new avoided level
structures 
are created (such terms were proposed to induce gaps otherwise forbidden in
magnetic molecules 
\cite{Barbara98}). This is a sufficient condition for having a finite
zero-field splitting with 
half-integer spins. The subsequent structure of energy spectrum at low energy
is generally like the 
one depicted in Fig.~\ref{figDM} for the case of a frustrated triangle (the 15
spins 1/2 form 3 
effective spins 1/2, antiferromagnetically coupled \cite{ChiorescuJMMM00}).
Here, we introduced the 
D-M interaction ($D=50$~mK) in a simple way to demonstrate the gap actually
opens with this type of 
perturbation. However, the real molecule consists of more complicated structure
where a spin of the 
inner triangle connects to spins of upper and lower hexagons of the V$_{15}$
molecule, and other 
D-M interactions can exist and should be taken into account
\cite{Konstantinidis02}. More realistic 
energy structure will 
be investigated after the determination of arrangement of D-M interactions on
the bonds 
\cite{DBCM}. For an effective triangle, there are two sets of avoided
structures and they cross in 
zero field, allowing the Kramers theorem to be satisfied. Depending on the
details of D-M interactions, 
two different situations may occur depending if the two sets of level repulsion
structures are or 
are not orthogonal to each other. In the first case the four states are
classified into two sets of 
subspaces, and the LZS mechanism exactly holds because the LZS transition
occurs independently in each 
branch of avoided level crossing structure. In the second case, the four states
form an 
irreductible space, where one of the state scatters into all of the four states
when the field is 
swept, and the LZS mechanism should not hold. However, even in this last case,
the change of 
magnetization at the crossing point as a function of the sweeping velocity is
almost perfectly 
expressed by the LZS formula \cite{Miyashita02}. As a consequence, one can say
that the change of 
magnetization in multispin systems with D-M interactions, should  follows the
LZS model in a wide 
range of interaction parameters.

In short, previous studies of the "butterfly hysteresis loop" in the multi-spin
V$_{15}$ molecule 
\cite{ChiorescuPRL00a,ChiorescuJMMM00} are now extended to the non-dissipative
LZS regime. First, 
we have shown that the crossover from the dissipative to the non-dissipative
LZS regime, can easily 
be achieved by changing the coupling of the sample to the cryostat. This first
demonstration of 
dissipation control in a two level system has important implications for
further applications on 
quantum information and computers. The role of the spin-bath 
becomes visible when experiments are performed near zero-field, in the level
mixing region. 
However, the associated relaxation times are still long enough to allow
adiabatic magnetization 
experiments. Secondly, we directly verified the zero-field gap previously
obtained in the 
dissipative regime and attributed to the effect of D-M interactions of this
multi-spin system. The 
third result is the clear comparison between experiments and theory, showing
that the conditions 
for the existence of a gap in a half-integer multi-spin system with
antisymmetric interactions 
(\emph{e.g.} D-M interaction) are fullfilled. 

\begin{acknowledgments}
We are very pleased to thank H. B\"ogge, A.K. Zvezdin, D. Mailly, K. Saito and
N. 
Nagaosa for on-going collaborations and useful discussions. 
\end{acknowledgments}

\bibliography{ic_nondissipative_tls}

\begin{thebibliography}{22}
\expandafter\ifx\csname natexlab\endcsname\relax\def\natexlab#1{#1}\fi
\expandafter\ifx\csname bibnamefont\endcsname\relax
  \def\bibnamefont#1{#1}\fi
\expandafter\ifx\csname bibfnamefont\endcsname\relax
  \def\bibfnamefont#1{#1}\fi
\expandafter\ifx\csname citenamefont\endcsname\relax
  \def\citenamefont#1{#1}\fi
\expandafter\ifx\csname url\endcsname\relax
  \def\url#1{\texttt{#1}}\fi
\expandafter\ifx\csname urlprefix\endcsname\relax\def\urlprefix{URL }\fi
\providecommand{\bibinfo}[2]{#2}
\providecommand{\eprint}[2][]{\url{#2}}

\bibitem[{\citenamefont{Leggett}(1980)}]{Leggett80}
\bibinfo{author}{\bibfnamefont{A.~J.} \bibnamefont{Leggett}},
  \bibinfo{journal}{Suppl. Progress of Theoretical Physics}
  \textbf{\bibinfo{volume}{69}}, \bibinfo{pages}{80} (\bibinfo{year}{1980}),
  \bibinfo{note}{{A}.~J.~Leggett and A.~Garg, Phys. Rev. Lett. \textbf{54}, 857
  (1985), A.~J.~Leggett, Science \textbf{296}, 861 (2002), M.~N.~Leuenberger
  and D. Loss, Nature \textbf{410}, 789 (2001)}.

\bibitem[{\citenamefont{Tupitsyn and Barbara}(2001)}]{Tupitsyn01}
\bibinfo{author}{\bibfnamefont{I.}~\bibnamefont{Tupitsyn}} \bibnamefont{and}
  \bibinfo{author}{\bibfnamefont{B.}~\bibnamefont{Barbara}}, in
  \emph{\bibinfo{booktitle}{Magneto-Science- From Molecules to Materials}},
  edited by \bibinfo{editor}{\bibfnamefont{M.}~\bibnamefont{Drillon}}
  \bibnamefont{and} \bibinfo{editor}{\bibfnamefont{J.}~\bibnamefont{Miller}}
  (\bibinfo{publisher}{Wiley VCH Verlag}, \bibinfo{year}{2001}).

\bibitem[{\citenamefont{Thomas et~al.}(1996)\citenamefont{Thomas, Lionti,
  Ballou, Gatteschi, Sessoli, and Barbara}}]{Thomas96}
\bibinfo{author}{\bibfnamefont{L.}~\bibnamefont{Thomas}},
  \bibinfo{author}{\bibfnamefont{F.}~\bibnamefont{Lionti}},
  \bibinfo{author}{\bibfnamefont{R.}~\bibnamefont{Ballou}},
  \bibinfo{author}{\bibfnamefont{D.}~\bibnamefont{Gatteschi}},
  \bibinfo{author}{\bibfnamefont{R.}~\bibnamefont{Sessoli}}, \bibnamefont{and}
  \bibinfo{author}{\bibfnamefont{B.}~\bibnamefont{Barbara}},
  \bibinfo{journal}{Nature} \textbf{\bibinfo{volume}{383}},
  \bibinfo{pages}{145} (\bibinfo{year}{1996}), \bibinfo{note}{{J}.~R.~Friedman,
  M.~P.~Sarachik, J.~Tejada and R.~Ziolo, Phys. Rev. Lett. \textbf{76}, 3830
  (1996), C. Sangregorio, T. Ohm, C. Paulsen, R. Sessoli and D. Gatteschi,
  Phys. Rev. Lett. \textbf{78}, 4645 (1997)}.

\bibitem[{\citenamefont{Wernsdorfer and Sessoli}(1999)}]{WW_Science99}
\bibinfo{author}{\bibfnamefont{W.}~\bibnamefont{Wernsdorfer}} \bibnamefont{and}
  \bibinfo{author}{\bibfnamefont{R.}~\bibnamefont{Sessoli}},
  \bibinfo{journal}{Science} \textbf{\bibinfo{volume}{284}},
  \bibinfo{pages}{133} (\bibinfo{year}{1999}).

\bibitem[{\citenamefont{Chiorescu
  et~al.}(2000{\natexlab{a}})\citenamefont{Chiorescu, Giraud, Jansen, Caneschi,
  and Barbara}}]{ChiorescuPRL00b}
\bibinfo{author}{\bibfnamefont{I.}~\bibnamefont{Chiorescu}},
  \bibinfo{author}{\bibfnamefont{R.}~\bibnamefont{Giraud}},
  \bibinfo{author}{\bibfnamefont{A.}~\bibnamefont{Jansen}},
  \bibinfo{author}{\bibfnamefont{A.}~\bibnamefont{Caneschi}}, \bibnamefont{and}
  \bibinfo{author}{\bibfnamefont{B.}~\bibnamefont{Barbara}},
  \bibinfo{journal}{Phys. Rev. Lett.} \textbf{\bibinfo{volume}{85}},
  \bibinfo{pages}{4807} (\bibinfo{year}{2000}{\natexlab{a}}),
  \bibinfo{note}{{F}. Luis, F.L. Mettes, J. Tejada, D. Gatteschi, L.J. de
  Jongh, Phys. Rev. Lett. \textbf{85}, 4377 (2000), G. Bellessa, N. Vernier, B.
  Barbara, D. Gatteschi, Phys. Rev. Lett. \textbf{83}, 416 (1999)}.

\bibitem[{\citenamefont{Chiorescu
  et~al.}(2000{\natexlab{b}})\citenamefont{Chiorescu, Wernsdorfer,
  M$\ddot{\mathrm{u}}$ller, B$\ddot{\mathrm{o}}$gge, and
  Barbara}}]{ChiorescuPRL00a}
\bibinfo{author}{\bibfnamefont{I.}~\bibnamefont{Chiorescu}},
  \bibinfo{author}{\bibfnamefont{W.}~\bibnamefont{Wernsdorfer}},
  \bibinfo{author}{\bibfnamefont{A.}~\bibnamefont{M$\ddot{\mathrm{u}}$ller}},
  \bibinfo{author}{\bibfnamefont{H.}~\bibnamefont{B$\ddot{\mathrm{o}}$gge}},
  \bibnamefont{and} \bibinfo{author}{\bibfnamefont{B.}~\bibnamefont{Barbara}},
  \bibinfo{journal}{Phys. Rev. Lett.} \textbf{\bibinfo{volume}{84}},
  \bibinfo{pages}{3454} (\bibinfo{year}{2000}{\natexlab{b}}).

\bibitem[{\citenamefont{Chiorescu
  et~al.}(2000{\natexlab{c}})\citenamefont{Chiorescu, Wernsdorfer, Barbara,
  M$\ddot{\mathrm{u}}$ller, and B$\ddot{\mathrm{o}}$gge}}]{ChiorescuJMMM00}
\bibinfo{author}{\bibfnamefont{I.}~\bibnamefont{Chiorescu}},
  \bibinfo{author}{\bibfnamefont{W.}~\bibnamefont{Wernsdorfer}},
  \bibinfo{author}{\bibfnamefont{B.}~\bibnamefont{Barbara}},
  \bibinfo{author}{\bibfnamefont{A.}~\bibnamefont{M$\ddot{\mathrm{u}}$ller}},
  \bibnamefont{and}
  \bibinfo{author}{\bibfnamefont{H.}~\bibnamefont{B$\ddot{\mathrm{o}}$gge}},
  \bibinfo{journal}{J. Magn. Magn. Mat.} \textbf{\bibinfo{volume}{{221(1-2)}}},
  \bibinfo{pages}{103} (\bibinfo{year}{2000}{\natexlab{c}}).

\bibitem[{\citenamefont{Barbara et~al.}(2002)\citenamefont{Barbara, Chiorescu,
  Wernsdorfer, B$\ddot{\mathrm{o}}$gge, and
  M$\ddot{\mathrm{u}}$ller}}]{Barbara02}
\bibinfo{author}{\bibfnamefont{B.}~\bibnamefont{Barbara}},
  \bibinfo{author}{\bibfnamefont{I.}~\bibnamefont{Chiorescu}},
  \bibinfo{author}{\bibfnamefont{W.}~\bibnamefont{Wernsdorfer}},
  \bibinfo{author}{\bibfnamefont{H.}~\bibnamefont{B$\ddot{\mathrm{o}}$gge}},
  \bibnamefont{and}
  \bibinfo{author}{\bibfnamefont{A.}~\bibnamefont{M$\ddot{\mathrm{u}}$ller}},
  \bibinfo{journal}{Progress Theor. Phys. Suppl.}
  \textbf{\bibinfo{volume}{145}}, \bibinfo{pages}{in press}
  (\bibinfo{year}{2002}).

\bibitem[{\citenamefont{Landau}(1932)}]{Landau32}
\bibinfo{author}{\bibfnamefont{L.}~\bibnamefont{Landau}},
  \bibinfo{journal}{Phys. Z. Sowjetunion} \textbf{\bibinfo{volume}{2}},
  \bibinfo{pages}{46} (\bibinfo{year}{1932}), \bibinfo{note}{{C}.~Zener, Proc.
  R. Soc. A \textbf{137}, 696 (1932), E.~C.~G. St$\ddot{\mathrm{u}}$ckelberg,
  Helv. Phys. Acta \textbf{5}, 369 (1932), S.~Miyashita, J. Phys. Soc. Jpn.
  \textbf{64}, 3207 (1995), V.V. Dobrovitski, A.K. Zvezdin, EuroPhys. Lett.
  \textbf{38}, 377 (1997).}

\bibitem[{\citenamefont{Leggett~et al}(1987)}]{Leggett87}
\bibinfo{author}{\bibfnamefont{A.}~\bibnamefont{Leggett~et al}},
  \bibinfo{journal}{Rev. Mod. Phys.} \textbf{\bibinfo{volume}{59}},
  \bibinfo{pages}{1} (\bibinfo{year}{1987}), \bibinfo{note}{{Y}. Gefen, E.
  Ben-Jacob, A. O. Caldeira, Phys. Rev. B \textbf{36}, 2770 (1987), Y.
  Kayanuma, H. Nakayama, Phys. Rev. B \textbf{57}, 13099 (1998), M. Grifoni, P.
  H$\ddot{\mathrm{a}}$nggi, Physics Reports \textbf{304}, 229 (1998).}

\bibitem[{\citenamefont{Schenker et~al.}()\citenamefont{Schenker, Leuenberger,
  Chaboussant, G$\ddot{\mathrm{u}}$del, and Loss}}]{Schenker02}
\bibinfo{author}{\bibfnamefont{R.}~\bibnamefont{Schenker}},
  \bibinfo{author}{\bibfnamefont{M.~N.} \bibnamefont{Leuenberger}},
  \bibinfo{author}{\bibfnamefont{G.}~\bibnamefont{Chaboussant}},
  \bibinfo{author}{\bibfnamefont{H.~U.} \bibnamefont{G$\ddot{\mathrm{u}}$del}},
  \bibnamefont{and} \bibinfo{author}{\bibfnamefont{D.}~\bibnamefont{Loss}},
  \bibinfo{note}{cond-mat/0204445. O. Waldmann, R. Koch, S. Schromm, P.
  M$\ddot{\mathrm{u}}$ller, I. Bernt, R. W. Saalfrank, cond-mat/0210309.}

\bibitem[{\citenamefont{Saito and Miyashita}(2001)}]{Saito01}
\bibinfo{author}{\bibfnamefont{K.}~\bibnamefont{Saito}} \bibnamefont{and}
  \bibinfo{author}{\bibfnamefont{S.}~\bibnamefont{Miyashita}},
  \bibinfo{journal}{J. Phys. Soc. Jpn.} \textbf{\bibinfo{volume}{70}},
  \bibinfo{pages}{3385} (\bibinfo{year}{2001}).

\bibitem[{\citenamefont{Nakano and Miyashita}(2001)}]{Nakano01}
\bibinfo{author}{\bibfnamefont{H.}~\bibnamefont{Nakano}} \bibnamefont{and}
  \bibinfo{author}{\bibfnamefont{S.}~\bibnamefont{Miyashita}},
  \bibinfo{journal}{J. Phys. Soc. Jpn.} \textbf{\bibinfo{volume}{70}},
  \bibinfo{pages}{2151} (\bibinfo{year}{2001}), \bibinfo{note}{{Y}. Ajiro and
  Y. Inagaki, private communications; Y. Narumi and K. Kindo, private
  communications.}

\bibitem[{\citenamefont{M$\ddot{\mathrm{u}}$ller and
  D$\ddot{\mathrm{o}}$ring}(1988)}]{Muller88}
\bibinfo{author}{\bibfnamefont{A.}~\bibnamefont{M$\ddot{\mathrm{u}}$ller}}
  \bibnamefont{and}
  \bibinfo{author}{\bibfnamefont{J.}~\bibnamefont{D$\ddot{\mathrm{o}}$ring}},
  \bibinfo{journal}{Angew. Chem., Intl. Ed. Engl.}
  \textbf{\bibinfo{volume}{27(12)}}, \bibinfo{pages}{1721}
  (\bibinfo{year}{1988}), \bibinfo{note}{{D.}~Gatteschi, L.~Pardi, A.-L. Barra,
  A.~M$\ddot{\mathrm{u}}$ller and J.~D$\ddot{\mathrm{o}}$ring, Nature
  \textbf{354}, 465 (1991).}

\bibitem[{\citenamefont{Abragam and Bleaney}(1970)}]{Abragam70}
\bibinfo{author}{\bibfnamefont{A.}~\bibnamefont{Abragam}} \bibnamefont{and}
  \bibinfo{author}{\bibfnamefont{B.}~\bibnamefont{Bleaney}},
  \emph{\bibinfo{title}{Electronic paramagnetic resonance of transition ions}}
  (\bibinfo{publisher}{Clarendon Press}, \bibinfo{address}{Oxford},
  \bibinfo{year}{1970}).

\bibitem[{\citenamefont{Prokof'ev and Stamp}(2000)}]{Prokofev00}
\bibinfo{author}{\bibfnamefont{N.}~\bibnamefont{Prokof'ev}} \bibnamefont{and}
  \bibinfo{author}{\bibfnamefont{P.}~\bibnamefont{Stamp}},
  \bibinfo{journal}{Rep. Prog. Phys.} \textbf{\bibinfo{volume}{63}},
  \bibinfo{pages}{669} (\bibinfo{year}{2000}).

\bibitem[{\citenamefont{Carter et~al.}(1977)\citenamefont{Carter, Bennett, and
  Kahan}}]{Carter77}
\bibinfo{author}{\bibfnamefont{G.}~\bibnamefont{Carter}},
  \bibinfo{author}{\bibfnamefont{L.}~\bibnamefont{Bennett}}, \bibnamefont{and}
  \bibinfo{author}{\bibfnamefont{D.}~\bibnamefont{Kahan}}, in
  \emph{\bibinfo{booktitle}{Metallic shifts in {NMR}}}, edited by
  \bibinfo{editor}{\bibfnamefont{B.}~\bibnamefont{Chalmers}},
  \bibinfo{editor}{\bibfnamefont{J.}~\bibnamefont{Christian}},
  \bibnamefont{and} \bibinfo{editor}{\bibfnamefont{T.}~\bibnamefont{Massalski}}
  (\bibinfo{publisher}{Pergamon Press}, \bibinfo{address}{London},
  \bibinfo{year}{1977}), vol. \bibinfo{volume}{20(I)} of
  \emph{\bibinfo{series}{Progress in materials science}}.

\bibitem[{\citenamefont{Chaboussant~et al}()}]{Chaboussant02}
\bibinfo{author}{\bibfnamefont{C.}~\bibnamefont{Chaboussant~et al}},
  \bibinfo{note}{cond-mat/0204365}.

\bibitem[{\citenamefont{Miyashita and Nagaosa}(2002)}]{Miyashita02}
\bibinfo{author}{\bibfnamefont{S.}~\bibnamefont{Miyashita}} \bibnamefont{and}
  \bibinfo{author}{\bibfnamefont{N.}~\bibnamefont{Nagaosa}},
  \bibinfo{journal}{Prog. Theor. Phys.} \textbf{\bibinfo{volume}{106}},
  \bibinfo{pages}{533} (\bibinfo{year}{2002}).

\bibitem[{\citenamefont{Barbara et~al.}(1998)\citenamefont{Barbara, Thomas,
  Lionti, Sulpice, and Caneschi}}]{Barbara98}
\bibinfo{author}{\bibfnamefont{B.}~\bibnamefont{Barbara}},
  \bibinfo{author}{\bibfnamefont{L.}~\bibnamefont{Thomas}},
  \bibinfo{author}{\bibfnamefont{F.}~\bibnamefont{Lionti}},
  \bibinfo{author}{\bibfnamefont{A.}~\bibnamefont{Sulpice}}, \bibnamefont{and}
  \bibinfo{author}{\bibfnamefont{A.}~\bibnamefont{Caneschi}},
  \bibinfo{journal}{J. Magn. Magn. Mat.} \textbf{\bibinfo{volume}{177-181}},
  \bibinfo{pages}{1324} (\bibinfo{year}{1998}).

\bibitem[{\citenamefont{Konstantinidis and Coffey}()}]{Konstantinidis02}
\bibinfo{author}{\bibfnamefont{N.~P.} \bibnamefont{Konstantinidis}}
  \bibnamefont{and} \bibinfo{author}{\bibfnamefont{D.}~\bibnamefont{Coffey}},
  \bibinfo{note}{cond-mat/0204435}.

\bibitem[{\citenamefont{de~Raedt et~al.}()\citenamefont{de~Raedt, Barbara,
  Chiorescu, and Miyashita}}]{DBCM}
\bibinfo{author}{\bibfnamefont{H.}~\bibnamefont{de~Raedt}},
  \bibinfo{author}{\bibfnamefont{B.}~\bibnamefont{Barbara}},
  \bibinfo{author}{\bibfnamefont{I.}~\bibnamefont{Chiorescu}},
  \bibnamefont{and}
  \bibinfo{author}{\bibfnamefont{S.}~\bibnamefont{Miyashita}},
  \bibinfo{note}{in preparation}.

\end{thebibliography}

\end{document}